\documentclass[twocolumn,showpacs,preprintnumbers,superscriptaddress,nofootinbib]{revtex4}
\usepackage{graphicx}

\def\lsim{\mathrel{\raise.3ex\hbox{$<$\kern-.75em\lower1ex\hbox{$\sim$}}}}
\def\gsim{\mathrel{\raise.3ex\hbox{$>$\kern-.75em\lower1ex\hbox{$\sim$}}}}

\begin{document}

\preprint{CALT-68-2392}
\preprint{FERMILAB-Pub-02/110-T}
\preprint{hep-ph/0206135}

\title{Variation of the cross section for $e^+e^- \to W^+H^-$ in the 
Minimal Supersymmetric Standard Model}

\author{Heather E. Logan}
\email{logan@pheno.physics.wisc.edu}
\affiliation{Theoretical Physics Department, Fermilab, PO Box 500, 
Batavia, Illinois 60510-0500, USA}
\affiliation{Department of Physics, University of Wisconsin,
Madison, Wisconsin 53706, USA} 

\author{Shufang Su}
\email{shufang@theory.caltech.edu}
\affiliation{California Institute of Technology, Pasadena, 
California 91125, USA}

\begin{abstract}
We study the loop-induced process $e^+e^- \to W^+H^-$ in the 
Minimal Supersymmetric Standard Model (MSSM).  
This process allows the charged Higgs boson to be produced in $e^+e^-$
collisions when $e^+e^- \to H^+H^-$ is kinematically forbidden due
to a large charged Higgs mass.
We scan over the MSSM parameters subject to experimental constraints
to examine the range of values the cross section can take.
We find that large enhancements of the cross section over that
in a non-supersymmetric two Higgs doublet model are possible,
especially for large $\tan\beta$ and light top and bottom squarks. 
Choosing a few typical MSSM parameter sets, we show the regions in the 
$m_{H^{\pm}}$--$\tan\beta$ plane in which at least 10 $W^{\pm}H^{\mp}$
events would be produced at the $e^+e^-$ collider.
\end{abstract}

\pacs{12.60.Jv, 12.60.Fr, 14.80.Cp, 14.80.Ly}

\maketitle


In a recent paper \cite{HW} we computed the cross section for 
$e^+e^- \to W^+H^-$ in the Minimal Supersymmetric Standard Model
(MSSM).  This process, which first arises at the one loop level, offers
the possibility of producing the charged Higgs boson 
with mass above half the $e^+e^-$ collider center-of-mass
energy $\sqrt{s}$, when $e^+e^- \to H^+H^-$ is kinematically forbidden.
The cross section for $e^+e^- \to W^+H^-$ is largest at
low values of $\tan\beta$,
and can be enhanced over its value in the 
non-supersymmetric two Higgs doublet model (2HDM) \cite{Arhrib,Zhu}
by the contributions of light supersymmetric (SUSY) 
particles \cite{HW}.  
This is the most promising process considered to 
date for producing charged Higgs bosons with mass above $\sqrt{s}/2$
in $e^+e^-$ collisions at low to moderate $\tan\beta$ values,
above the lower bound of $\tan\beta \geq 2.4$ from the CERN 
$e^+e^-$ collider LEP-2 
Higgs search \cite{LEP2}.
This region of low to moderate $\tan\beta$ is 
exactly where the CERN Large Hadron Collider (LHC) will have difficulty 
discovering the heavy MSSM Higgs bosons \cite{AtlasTDR,AtlasH+,CMS}.

The goal of this paper is to examine the range of the 
$e^+e^- \to W^+H^-$ cross section
over the MSSM parameter space, taking into account the present
experimental constraints on SUSY particle masses.

The strongest constraint on the MSSM Higgs sector 
is the lower bound on the mass of the lighter
CP-even Higgs boson $h^0$ from searches at LEP \cite{LEP2}.  
For given values of $m_{H^{\pm}}$ and $\tan\beta$,
the MSSM prediction for $m_{h^0}$ receives very large radiative 
corrections of several tens of GeV \cite{HiggsRCs}, primarily due to loops
involving top quarks and their supersymmetric partners.
Obviously, these radiative corrections must be taken into account if the 
experimental constraint on $m_{h^0}$ is to be used to constrain the 
MSSM parameter space, and we do so here.  

The radiative corrections to the MSSM Higgs sector
also appear directly in the calculation of the 
$e^+e^- \to W^+H^-$ cross section via the dependence of 
Feynman diagrams \cite{HW,Arhrib} with Higgs bosons in the loop
on $m_{h^0}$ and the effective mixing angle $\alpha_{\rm eff}$, 
which diagonalizes the CP-even Higgs mass eigenstates and affects 
the Higgs couplings \cite{HHG}.
Formally, these corrections 
are of two-loop and higher orders in the cross section; 
however, because their effects can be large (especially on $m_{h^0}$), 
we have examined their impact using the program FeynHiggs 
\cite{FeynHiggs}, which
includes the complete one-loop and leading two-loop radiative corrections
to the MSSM Higgs boson masses and mixing angle calculated in the 
Feynman-diagrammatic approach.
In fact, we find that the effect
on the cross section is quite small, about 5\% or less.
Nevertheless, we have included these radiative corrections in the 
numerical results throughout this paper.


We now explore the possible variation of the 
$e^+e^- \to W^+H^-$ cross section as a function of the MSSM parameters.
The SUSY contributions to this cross section come mainly from loop diagrams
involving top and bottom squarks and, to a lesser extent, 
charginos and neutralinos.
The diagrams involving top and bottom squarks are enhanced by the 
large third-generation Yukawa couplings, but their contribution decouples
quickly, like $m_{\tilde t,\tilde b}^{-2}$, as their masses 
$m_{\tilde t, \tilde b}$ increase.
The diagrams involving charginos and neutralinos decouple more slowly, 
like $m_{\tilde \chi}^{-1}$, as the relevant chargino or neutralino 
mass $m_{\tilde \chi}$ increases.
There are additional contributions from box and $t$-channel diagrams involving
charginos, neutralinos, selectrons, and/or sneutrinos, 
and from diagrams involving loops of sleptons and of the first two 
generations of squarks that couple to the charged Higgs boson through 
the supersymmetric D-terms;
however, the contributions of these diagrams are small in general.
For explicit expressions, see Ref.~\cite{HW}.

\begin{table}
\begin{tabular}{|c|cc|}
\hline
Parameter & Min & Max \\
\hline
$\tan\beta$ & 2.4 & 60 \\
$M_{\rm SUSY}^{tb}$ & 50 GeV & 1000 GeV \\
$M_{\rm SUSY}$ & 50 GeV & 1000 GeV \\
$|\mu|$ & 50 GeV & 1000 GeV \\
$M_1$ & 50 GeV & 1000 GeV \\
$M_2$ & 50 GeV & 1000 GeV \\
$m_{H^{\pm}}$ & $\sqrt{s}/2$ & $\sqrt{s} - 100$ GeV \\
\hline
\end{tabular}
\caption{Ranges of SUSY parameters scanned.  The $\mu$ parameter can
take either sign.}
\label{tab:scanparams}
\end{table}

We begin by scanning over the MSSM parameters in the ranges shown 
in Table~\ref{tab:scanparams}.  We use a linear metric for all
parameters except $\tan\beta$, for which we use a logarithmic metric.
We vary separately the soft SUSY breaking mass $M_{\rm SUSY}^{tb}$ of
the top and bottom squarks and the mass $M_{\rm SUSY}$ of the 
sleptons and first two generations of squarks in order to be able to 
separate the effects of the top and bottom squarks.\footnote{A large
hierarchy between the masses of the third generation squarks and the first
two generations can lead to dangerous flavor changing neutral currents
in the $B$ sector only if there are off-diagonal squark mass terms that mix 
generations.  This is strongly dependent on the flavor structure of
the model and we do not consider it here.}
We take $A_t = A_b = 2 M_{\rm SUSY}^{tb}$, which yields the maximal 
value for $m_{h^0}$.
The gluino mass enters our calculation only through the two-loop radiative 
corrections to the Higgs sector and has almost no effect on the cross 
section; here we set it to 800 GeV.

We discard any points that yield SUSY particle masses below the
experimental lower bounds summarized in Table~\ref{tab:SUSYbounds}.
These bounds are model dependent; we use the bounds quoted in
Ref.~\cite{ShufangHiggs} for the minimal supergravity scenario, which is 
reasonable for a large fraction of the scanned parameter space.
\begin{table}
\begin{tabular}{|c|c|c|c|c|c|c|}
\hline
$\tilde \nu$ & $\tilde \ell$ & $\tilde q$
& $\tilde t_1$ & $\tilde b_1$
  & $\tilde \chi_1^0$ 
& $\tilde \chi_1^{\pm}$ \\
\hline
43 GeV       & 95 GeV  & 100 GeV         & 95 GeV       & 85 GeV
  & 36 GeV            & 84.6 GeV              \\
\hline
\end{tabular}
\caption{Lower bounds imposed on the SUSY particle masses.  $\tilde q$ denotes
squarks of the first two generations.}
\label{tab:SUSYbounds}
\end{table}
We also discard any points that give too small a prediction for
$m_{h^0}$.  
For the large $m_{H^{\pm}}$ values that we consider,
the LEP bound on $m_{h^0}$ coincides with the 
Standard Model Higgs mass bound of about 114 GeV \cite{LEPHiggsbound}.
After the dominant two-loop radiative corrections to $m_{h^0}$ \cite{FeynHiggs}
have been included, 
the remaining theoretical uncertainty due to uncalculated 
higher order corrections is estimated to be about $\pm 3$ GeV \cite{Sven}.
We take this into account by lowering the LEP bound by 3 GeV; 
{\it i.e.}, we require $m_{h^0} > 111$ GeV.
Finally, we discard any points that give too large a SUSY contribution to the
$\rho$ parameter \cite{rhoSUSY}; we require 
$|\Delta \rho_{\rm SUSY}| < 0.002$ \cite{PDG}.\footnote{In principle, 
one could also impose the constraints 
from the measurement of $b \to s \gamma$, which receives MSSM contributions
from diagrams involving charged Higgs boson or chargino exchange, especially
at large $\tan\beta$.  However, because $b \to s \gamma$ is
a flavor-changing decay, the constraints that it places on the MSSM
parameters depend strongly on the presence or absence of 
additional non-minimal flavor structure in the model.  
Similarly, one could discard any points that yield an unacceptable amount
of neutralino dark matter.  However, introducing a very small amount of 
$R$-parity violation can remove the dark matter constraint
while having a negligible effect on the collider phenomenology.
In order to maintain generality, then, we have not applied the 
$b \to s \gamma$ or dark matter constraints.}

\begin{figure*}
\resizebox{17cm}{!}{
\includegraphics{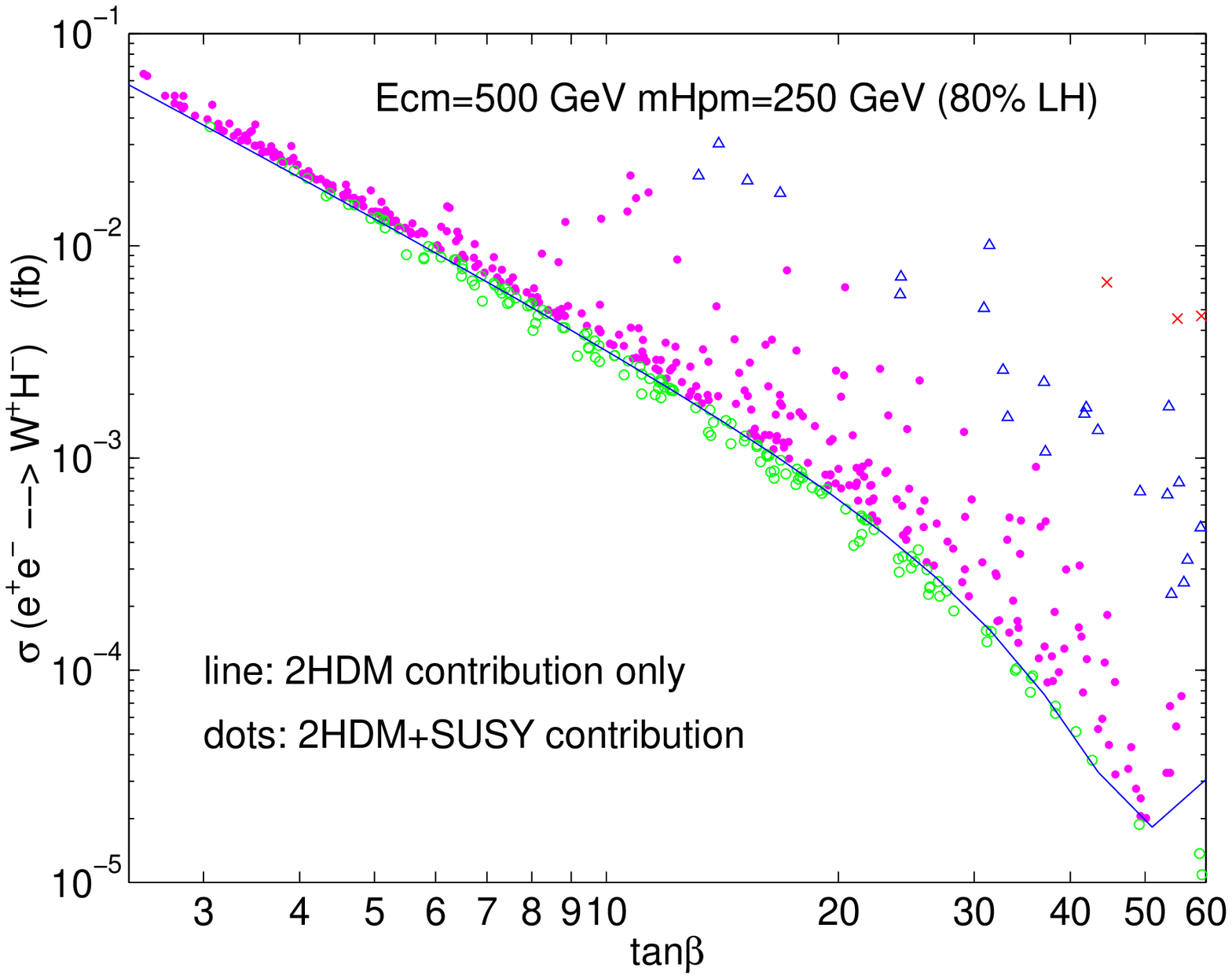}
\includegraphics{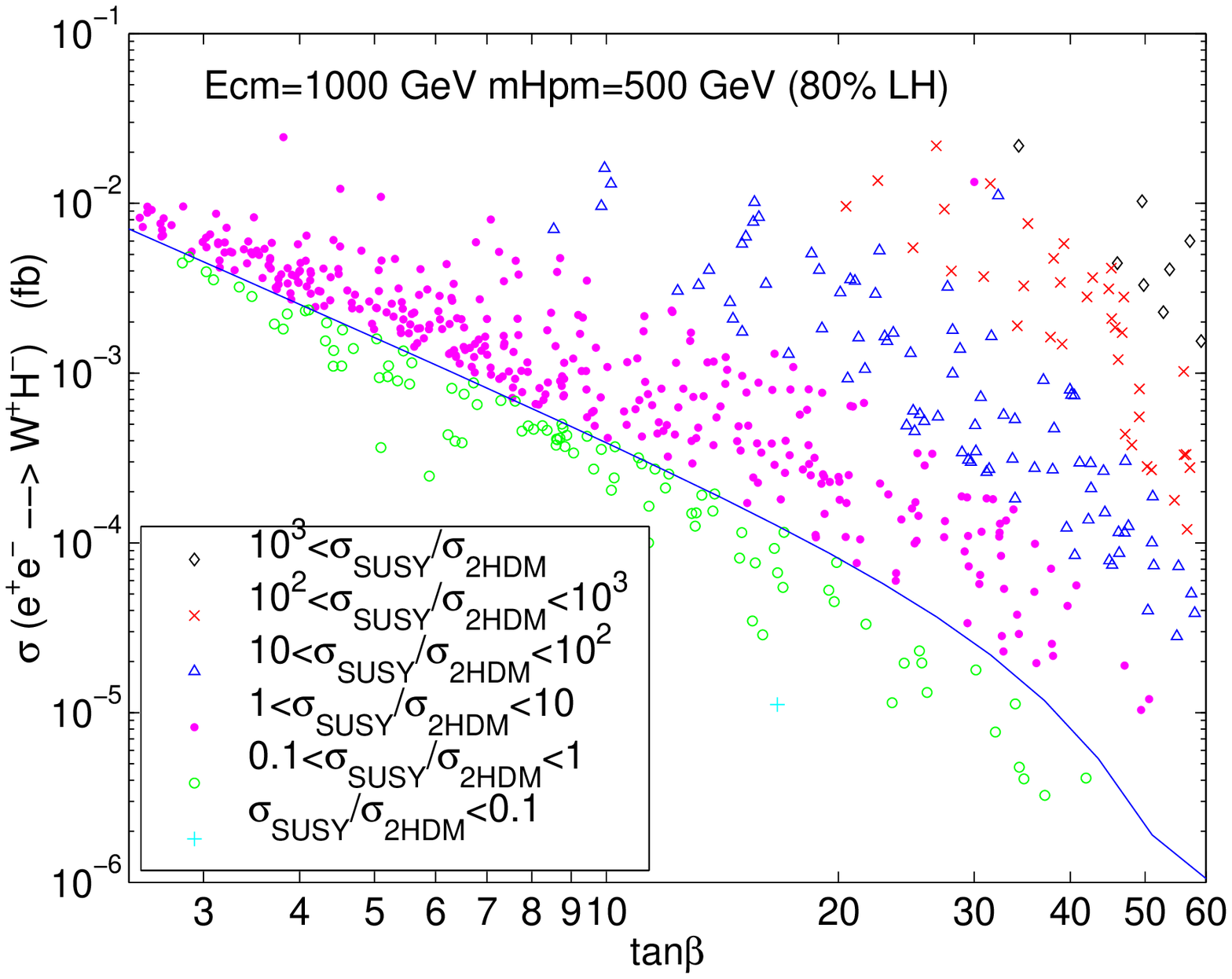}}
\caption{MSSM cross section as a function of $\tan\beta$ for
$\sqrt{s} = 500$ GeV and $m_{H^{\pm}} = 250$ GeV (left) 
and $\sqrt{s} = 1000$ GeV and $m_{H^{\pm}} = 500$ GeV (right).
The different symbols show the enhancement of the MSSM cross section
relative to the 2HDM.
The solid line shows the 2HDM cross section.
}
\label{fig:tanbeta}
\end{figure*}

\begin{figure}
\resizebox{8.5cm}{!}{
\includegraphics{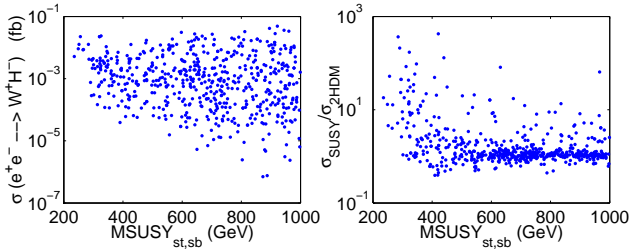}}
\caption{MSSM cross section (left) and enhancement relative to
the 2HDM (right) as a function of $M_{\rm SUSY}^{tb}$, 
for $\sqrt{s} = 500$ GeV.
}
\label{fig:mst}
\end{figure}

\begin{figure}
\resizebox{8.5cm}{!}{
\includegraphics{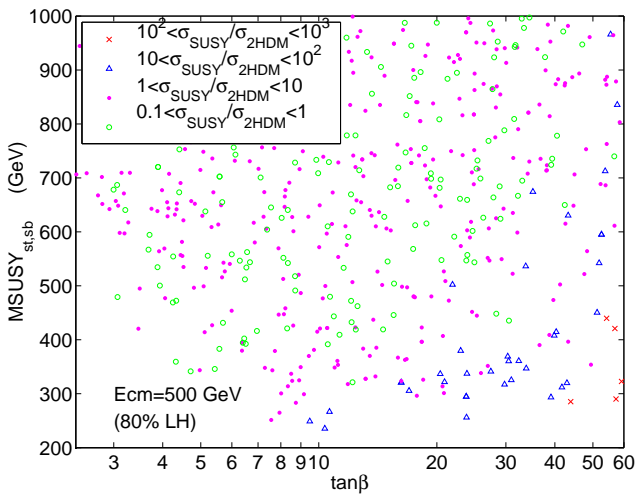}}
\caption{The MSSM enhancement of the cross section relative to its 2HDM
value as a function of $M_{\rm SUSY}^{tb}$ and $\tan\beta$, for
$\sqrt{s} = 500$ GeV.}
\label{fig:beta_mst}
\end{figure}

The results of the parameter scans are shown in 
Figs.~\ref{fig:tanbeta}-\ref{fig:beta_mst}.  
We assume 80\% left-polarized electron beams and unpolarized positrons.
Where the MSSM cross sections are compared to the 2HDM, we use the MSSM
relations for the Higgs masses and mixing angle including 
radiative corrections
in the 2HDM calculation in order to isolate the effects of the SUSY 
loop diagrams.

In Fig.~\ref{fig:tanbeta} we compare the cross section in the MSSM with that
in the 2HDM as a function of $\tan\beta$, 
for $\sqrt{s} = 500$ and 1000 GeV.  In 
Fig.~\ref{fig:tanbeta} only we fix $m_{H^{\pm}}$ to a single value, 
$m_{H^{\pm}} = \sqrt{s}/2$, in order to more clearly illustrate the 
effect of the MSSM contributions.
While the 2HDM cross section falls rapidly with increasing $\tan\beta$,
the MSSM contributions depend much more weakly on $\tan\beta$,
especially at $\sqrt{s} = 1000$ GeV, where the maximum cross section is
almost independent of $\tan\beta$.  This implies that the
largest relative cross section enhancements due to MSSM contributions 
occur at large $\tan\beta$, as shown by the different symbols in 
Fig.~\ref{fig:tanbeta}.  At a 500 GeV machine, an enhancement of more than
a factor of 10 can occur for $\tan\beta > 10$, while a factor of
100 can occur for $\tan\beta > 40$.  At a 1000 GeV machine,
the enhancements can be even larger.
At low $\tan\beta$, cross section enhancements of roughly 50\% are typical. 

Assuming an integrated luminosity of 500 fb$^{-1}$ at $\sqrt{s} = 500$ GeV
(1000 fb$^{-1}$ at $\sqrt{s} = 1000$ GeV), a cross section above
0.01 fb (0.005 fb) will yield 
at least 10 $W^{\pm}H^{\mp}$ events.\footnote{We choose this 10-event
criterion as a measure of the relevance of the $e^+e^- \to W^+H^-$ 
production process.  We expect that this is too few events to 
allow a discovery; a background study is needed to establish the 
minimum observable signal cross section.}
Fig.~\ref{fig:tanbeta} shows that these cross sections can be reached
for most values of $\tan\beta$ thanks to large MSSM enhancements, 
especially at $\sqrt{s} = 1000$ GeV.

In Fig.~\ref{fig:mst} we show the cross section (left) 
and the enhancement relative to the 2HDM (right)
as a function of $M_{\rm SUSY}^{tb}$, for $\sqrt{s} = 500$ GeV.
The maximum cross section is roughly independent of $M_{\rm SUSY}^{tb}$,
while the minimum cross section decreases with increasing $M_{\rm SUSY}^{tb}$.
The maximum cross section occurs mainly at small $m_{H^{\pm}}$ and 
low $\tan\beta$, for which the 2HDM cross section is already large;
SUSY loop contributions yield only a moderate enhancement in this regime.
The minimum cross section typically occurs at large $\tan\beta$
when the 2HDM cross section is very small; in this regime
large enhancements of the cross section due to top and bottom squark loops
occur for low $M_{\rm SUSY}^{tb}$ values, below about 500 GeV.
This joint dependence of the enhancement on $\tan\beta$ and 
$M_{\rm SUSY}^{tb}$ is illustrated in Fig.~\ref{fig:beta_mst}, which shows
that large enhancements relative to the 2HDM
(triangles and crosses in Fig.~\ref{fig:beta_mst})
require both relatively low $M_{\rm SUSY}^{tb}$ and large $\tan\beta$.
The situation is similar at $\sqrt{s} = 1000$ GeV.

Finally, the dependence of the cross section on the remaining
parameters, $M_{\rm SUSY}$, $\mu$, $M_1$ and $M_2$, is weaker than 
the $M_{\rm SUSY}^{tb}$ dependence, indicating
that the contributions of diagrams involving charginos, neutralinos,
sleptons and/or the first two generations of squarks are generally less
important than the top and bottom squark contributions.

\begin{figure*}
\resizebox{17cm}{!}{
\includegraphics{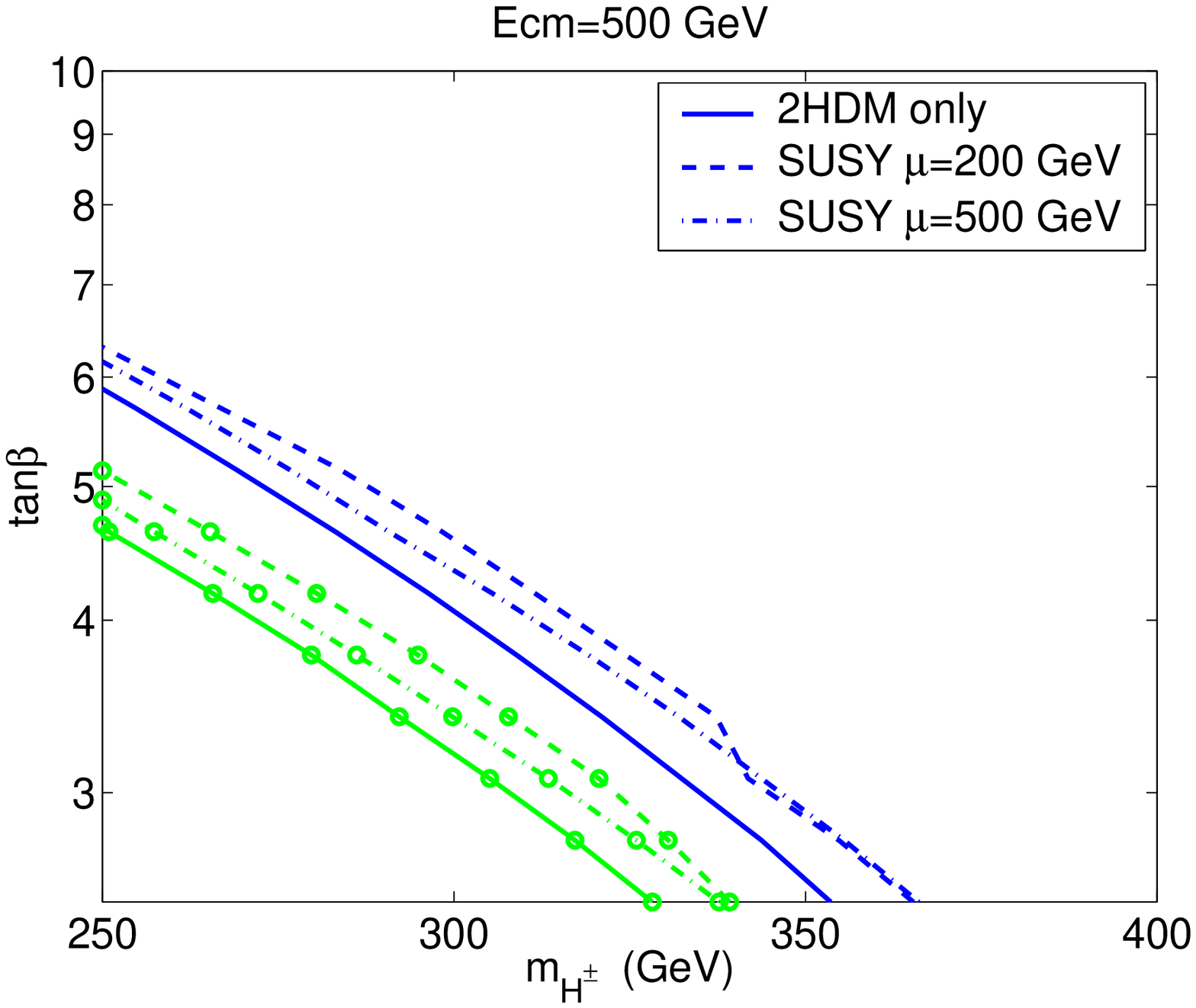}
\includegraphics{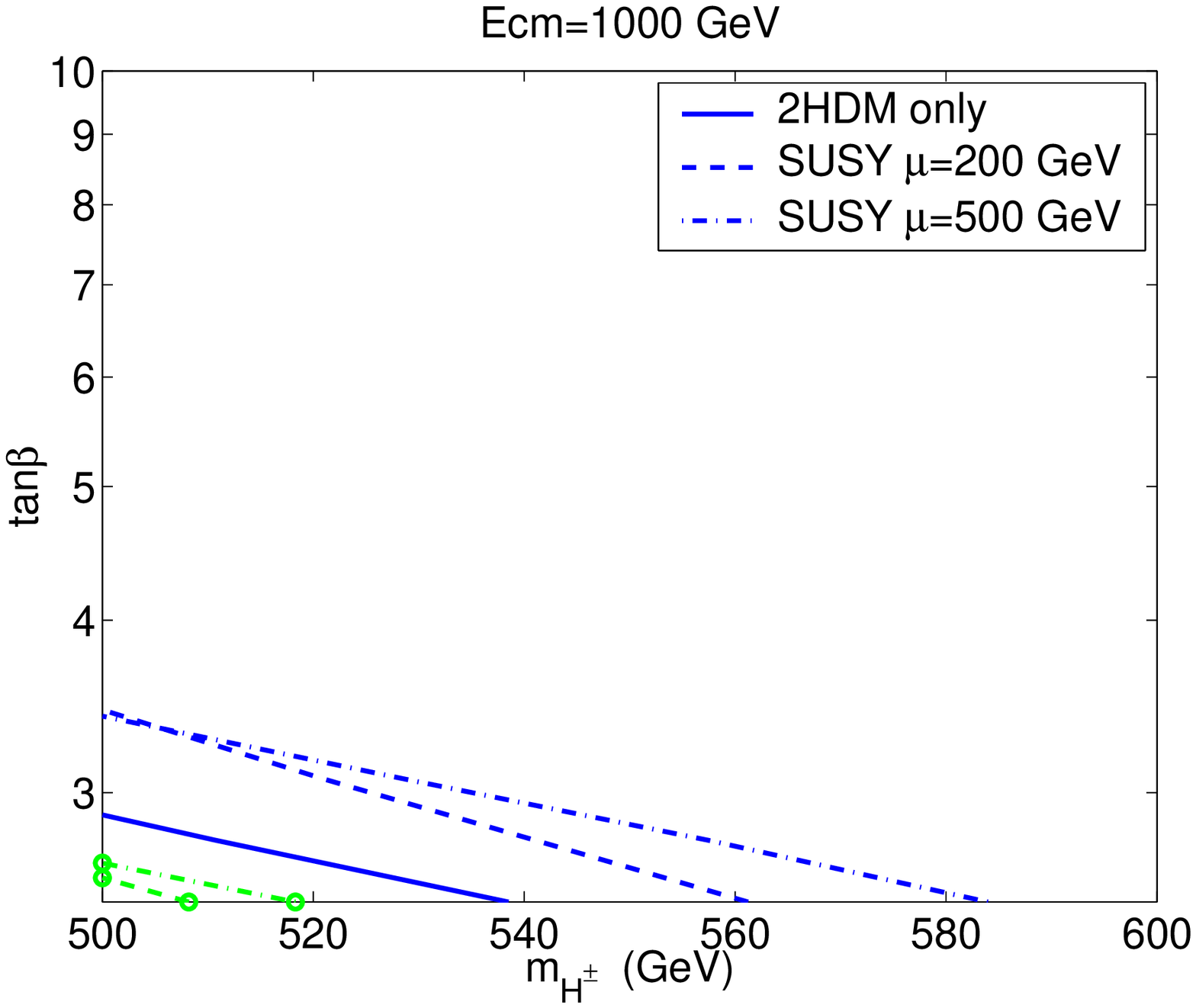}}
\caption{Ten-event contours for $e^+e^- \to W^{\pm}H^{\mp}$
for $\sqrt{s} = 500$ GeV with 500 fb$^{-1}$ (left)
and $\sqrt{s} = 1000$ GeV with 1000 fb$^{-1}$ (right).
The SUSY parameters are chosen to be $M_{\rm SUSY}^{tb} = 1000$ GeV for the
top and bottom squarks, $M_{\rm SUSY} = 200$ GeV for the rest of the squarks
and the sleptons, $2M_1 = M_2 = 200$ GeV, $M_{\tilde g} = 800$ GeV,
$A_t = A_b = 2 M_{\rm SUSY}^{tb}$,
and $\mu = 200$ GeV (dashed lines) and 500 GeV (dot-dashed lines).
Solid lines show the 2HDM result. 
Light (green) lines show the unpolarized cross section and dark (blue) lines
show the cross section with an 80\% left-polarized $e^-$ beam.
}
\label{fig:contours}
\end{figure*}
In order to illustrate the reach of the $e^+e^- \to W^+H^-$ production
process, we choose a ``typical'' set of SUSY parameters with two values
for the $\mu$ parameter (see the caption of
Fig.~\ref{fig:contours}) that obeys the experimental constraints discussed
in the previous section.
For this set of parameters, we show contours 
in Fig.~\ref{fig:contours} in the $m_{H^{\pm}}$--$\tan\beta$ plane below 
which 10 or more $W^{\pm}H^{\mp}$ events will be produced 
in the $e^+e^-$ collider data sample.
As a typical collider run plan, we 
assume a final integrated luminosity of 500 fb$^{-1}$ at $\sqrt{s} = 500$ GeV 
and 1000 fb$^{-1}$ at $\sqrt{s} = 1000$ GeV.
We plot 10-event contours for the 2HDM\footnote{The radiative corrections to 
the MSSM Higgs boson masses and mixing angle have been included for both 
the 2HDM and MSSM contours.  For the 2HDM Higgs sector 
radiative corrections we take $\mu = 200$ GeV; taking $\mu = 500$ GeV 
changes the cross section by less than 1\%.} 
(solid lines) and for the MSSM with $\mu = 200$ GeV (dashed lines) 
and 500 GeV (dot-dashed lines).
We consider an unpolarized $e^-$ beam
(light or green lines) and an 80\% left-polarized $e^-$ beam 
(dark or blue lines).
In all cases we assume the $e^+$ beam is unpolarized.

The maximum 10-event reaches in $m_{H^{\pm}}$ and $\tan\beta$ 
shown in Fig.~\ref{fig:contours} are summarized in Table~\ref{tab:reaches}.
In all cases considered, the MSSM contributions increase the 10-event 
reach over that in the 2HDM:
the reach in $m_{H^{\pm}}$ is increased by about 10 GeV at 
$\sqrt{s} = 500$ GeV, and by about 20 GeV or more at $\sqrt{s} = 1000$ GeV.
The $\mu$ dependence is fairly weak.
Using an 80\% left-polarized $e^-$ beam increases the reach compared to
using unpolarized beams by about 30 GeV in $m_{H^{\pm}}$ at 
$\sqrt{s} = 500$ GeV and by twice that at $\sqrt{s} = 1000$ GeV; at either
center-of-mass energy the reach in $\tan\beta$ increases by about 1 unit.
\begin{table}
\begin{tabular}{|c|cc|cc|}
\hline
 & \multicolumn{2}{c|}{unpolarized} & \multicolumn{2}{c|}{polarized} \\
$\sqrt{s} = 500$ GeV & $m_{H^{\pm}}$ & $\tan\beta$ & 
	$m_{H^{\pm}}$ & $\tan\beta$ \\
\hline
2HDM & 328 GeV & 4.7 & 354 GeV & 5.9 \\
MSSM, $\mu = 200$ GeV & 339 GeV & 5.1 & 365 GeV & 6.3 \\
MSSM, $\mu = 500$ GeV & 338 GeV & 4.9 & 366 GeV & 6.2 \\
\hline
$\sqrt{s} = 1000$ GeV & & & & \\
\hline
2HDM & -- & -- & 539 GeV & 2.9 \\
MSSM, $\mu = 200$ GeV & 508 GeV & 2.6 & 561 GeV & 3.4 \\
MSSM, $\mu = 500$ GeV & 518 GeV & 2.7 & 584 GeV & 3.4 \\
\hline 
\end{tabular}
\caption{Maximum 10-event 
reach in $m_{H^{\pm}}$ and $\tan\beta$ for the 2HDM and
MSSM, from Fig.~\ref{fig:contours}.}
\label{tab:reaches}
\end{table}


To summarize, we examined the range of values for the cross 
section for $e^+e^- \to W^+H^-$
possible in the MSSM by scanning over the MSSM parameters, imposing the 
experimental constraints on the $\rho$ parameter and the
masses of the SUSY particles and the lighter CP-even Higgs boson $h^0$.
While in the 2HDM the cross section falls rapidly with increasing 
$\tan\beta$, the maximal cross section values in the MSSM are roughly
independent of $\tan\beta$, especially at higher collider center-of-mass
energies.  In particular, very large enhancements of the cross 
section relative to its value in the 2HDM are possible at
large $\tan\beta$.  These large enhancements at large $\tan\beta$ 
typically occur for low top and bottom squark masses.
For low $\tan\beta$, where the 2HDM cross section reaches its maximum
value, enhancements of the cross section by about 50\% in the MSSM 
are typical.

We also examined in detail the reach in $m_{H^{\pm}}$ and $\tan\beta$,
focusing on the dependence on the $\mu$ parameter and electron beam
polarization.
Because the cross section is quite sensitive to the SUSY parameters,
this process can be used not only to produce $H^{\pm}$ at low $\tan\beta$,
but also to test and/or constrain the MSSM \cite{Hprodobs}.

Finally, we found that using the radiatively corrected
MSSM Higgs boson masses and mixing angle in the cross section calculation 
instead of the tree-level masses and mixing angle has only a small numerical
effect on the cross section.

\vskip0.5cm
\noindent
{\it Note added:} As we were finishing this paper, we learned about
another group \cite{Breinee} working on the same subject.

\begin{acknowledgments}
We thank Sven Heinemeyer for many helpful discussions and suggestions.
We thank Oliver Brein for comparison of the numerical results. 
Fermilab is operated by Universities Research Association Inc.\
under contract no.~DE-AC02-76CH03000 with the U.S. Department of
Energy.  S.S. is supported by the DOE under grant DE-FG03-92-ER-40701 and 
by the John A. McCone Fellowship.
\end{acknowledgments}



\begin{thebibliography}{99}

\bibitem{HW}
H.~E.~Logan and S.~Su,
Phys.\ Rev.\ D {\bf 66}, 035001 (2002).

\bibitem{Arhrib}
A.~Arhrib, M.~Capdequi Peyranere, W.~Hollik and G.~Moultaka,
Nucl.\ Phys.\ B {\bf 581}, 34 (2000);
S.~Kanemura,
Eur.\ Phys.\ J.\ C {\bf 17}, 473 (2000).

\bibitem{Zhu}
S.~H.~Zhu,
arXiv:hep-ph/9901221.

\bibitem{LEP2}
LEP Higgs Working Group Collaboration,
arXiv:hep-ex/0107030.

\bibitem{AtlasTDR}
K.~Lassila-Perini, ETH Dissertation thesis No. 12961 (1998);
ATLAS collaboration Technical Design Report, available from
\verb+http://atlasinfo.cern.ch/Atlas/+
\verb+GROUPS/PHYSICS/TDR/access.html+.

\bibitem{AtlasH+}
K.~A.~Assamagan, Y.~Coadou and A.~Deandrea,
EPJdirect C {\bf 09}, 1 (2002);

K.~A.~Assamagan and Y.~Coadou,
Acta Phys.\ Polon.\ B {\bf 33}, 707 (2002).

\bibitem{CMS}
D.~Denegri {\it et al.},
arXiv:hep-ph/0112045.

\bibitem{HiggsRCs}
For a review and references, see 
M.~Carena, H.~E.~Haber, S.~Heinemeyer, W.~Hollik, C.~E.~Wagner and G.~Weiglein,
Nucl.\ Phys.\ B {\bf 580}, 29 (2000).

\bibitem{HHG}
J.~F.~Gunion and H.~E.~Haber,
Nucl.\ Phys.\ B {\bf 272}, 1 (1986)
[Erratum-ibid.\ B {\bf 402}, 567 (1993)];
Nucl.\ Phys.\ B {\bf 278}, 449 (1986);
J.~F.~Gunion, H.~E.~Haber, G.~L.~Kane and S.~Dawson,
{\it The Higgs Hunter's Guide,}
(Perseus Publishing, Cambridge, MA, 2000).

\bibitem{FeynHiggs}
S.~Heinemeyer, W.~Hollik and G.~Weiglein,
Comput.\ Phys.\ Commun.\  {\bf 124}, 76 (2000);
see \verb+http://www.feynhiggs.de+.

\bibitem{ShufangHiggs}
S.~Ambrosanio, A.~Dedes, S.~Heinemeyer, S.~Su and G.~Weiglein,
Nucl.\ Phys.\ B {\bf 624}, 3 (2002).

\bibitem{LEPHiggsbound}
R.~Barate {\it et al.}  [ALEPH Collaboration],
Phys.\ Lett.\ B {\bf 499}, 53 (2001);
LEP Higgs Working Group,
arXiv:hep-ex/0107029 (LHWG/2001-03).

\bibitem{Sven}
S.~Heinemeyer, W.~Hollik and G.~Weiglein,
Eur.\ Phys.\ J.\ C {\bf 9}, 343 (1999).

\bibitem{rhoSUSY}
A.~Djouadi, P.~Gambino, S.~Heinemeyer, W.~Hollik, C.~Junger and G.~Weiglein,
Phys.\ Rev.\ D {\bf 57}, 4179 (1998).

\bibitem{PDG}
D.~E.~Groom {\it et al.}  [Particle Data Group Collaboration],
Eur.\ Phys.\ J.\ C {\bf 15}, 1 (2000),
and 2001 partial update for 2002 edition (URL: \verb+http://pdg.lbl.gov+).

\bibitem{Hprodobs}
The use of a Higgs boson production cross section to constrain
the MSSM has also been considered in
S.~Dawson and S.~Heinemeyer,
Phys.\ Rev.\ D {\bf 66}, 055002 (2002).

\bibitem{Breinee}
O.~Brein, W.~Hollik and T.~Hahn, in preparation;
O.~Brein,
arXiv:hep-ph/0209124.

\end{thebibliography}
\end{document}